\documentclass[namedreferences]{SolarPhysics_arxiv}
\usepackage[optionalrh]{spr-sola-addons_arxiv} % For arxiv
\usepackage{graphicx}        % For eps figures, newer & more powerfull
\usepackage{color}           % For color text: \color command
\usepackage{url}             % For breaking URLs easily trough lines
\newcommand{\Sunrise}{{\it Sunrise }}

\newcommand{\aap}{    {\it Astron. Astrophys.}}
\newcommand{\aaps}{   {\it Astron. Astrophys. Suppl.}}

\newcommand{\apj}{    {\it Astrophys. J.}}

\begin{document}

\begin{article}

\begin{opening}

%%%%%%%%%%%%%%%%%%%%%%%%%%%%%%%%%%%%%%%%
%

%
   \title{The Filter Imager SuFI and the Image Stabilization and Light Distribution System ISLiD of
   the \Sunrise Balloon-Borne Observatory: Instrument Description}

%\titlerunning{SuFI and ISLiD onboard \Sunrise}

\author{A.~\surname{Gandorfer}$^{1}$\sep
        B.~\surname{Grauf}$^{1}$\sep
        P.~\surname{Barthol}$^{1}$\sep
        T.L.~\surname{Riethm\"{u}ller}$^{1}$\sep
        S.K.~\surname{Solanki}$^{1}$\sep
        B.~\surname{Chares}$^{1}$\sep
        W.~\surname{Deutsch}$^{1}$\sep
        S.~\surname{Ebert}$^{1}$\thanks{Now at EMCO MAGDEBURG AG,
Gr\"{o}perstra\ss e 18, D-39124 Magdeburg, Germany}\sep
        A.~\surname{Feller}$^{1}$\sep
        D.~\surname{Germerott}$^{1}$\sep
        K.~\surname{Heerlein}$^{1}$\sep
        J.~\surname{Heinrichs}$^{1}$\sep
        D.~\surname{Hirche}$^{1}$\sep
        J.~\surname{Hirzberger}$^{1}$\sep
        M.~\surname{Kolleck}$^{1}$\sep
        R.~\surname{Meller}$^{1}$\sep
        R.~\surname{M\"{u}ller}$^{1}$\sep
        R.~\surname{Sch\"{a}fer}$^{1}$\sep
        G.~\surname{Tomasch}$^{1}$\sep
        M.~\surname{Kn\"{o}lker}$^{2}$\sep
        V.~\surname{Mart\'{i}nez Pillet}$^{3}$\sep
        J.A.~\surname{Bonet}$^{3}$\sep
        W.~\surname{Schmidt}$^{4}$\sep
        T.~\surname{Berkefeld}$^{4}$\sep
        B.~\surname{Feger}$^{4}$\sep
        F.~\surname{Heidecke}$^{4}$\sep
        D.~\surname{Soltau}$^{4}$\sep
        A.~\surname{Tischenberg}$^{4}$\sep
        A.~\surname{Fischer}$^{4}$\sep
        A.~\surname{Title}$^{5}$\sep
        H.~\surname{Anwand}$^{1}$\thanks{Now at Georg-August-Universit\"{a}t, Institut f\"{u}r Astrophysik, Friedrich-Hund-Platz
            1, D-37077 G\"{o}ttingen, Germany}\sep
        E.~\surname{Schmidt}$^{6}$
      }
\runningauthor{Gandorfer et al.}
\runningtitle{SuFI and ISLiD onboard \Sunrise}

   \institute{$^{1}$ Max-Planck-Institut f\"{u}r Sonnensystemforschung,
  Max-Planck-Stra\ss e 2, D-37191 Katlenburg-Lindau, Germany\\
                     email: \url{gandorfer@mps.mpg.de} \\
              $^{2}$ High Altitude Observatory\thanks{HAO/NCAR is sponsored by the National Science Foundation}, P.O. Box 3000, Colorado 80301,
              USA\\
%                     email: \url{mknoelker@hao.ucar.edu} \\
              $^{3}$ Instituto de Astrof\'{i}sica de Canarias,
                        C/ Via L\'{a}ctea, s/n,
                              E38205 - La Laguna (Tenerife), Spain \\
%                     email: \url{vmp@iac.es} \\
              $^{4}$ Kiepenheuer-Institut f\"{u}r Sonnenphysik, Sch\"{o}neckstra\ss e 6, D-79104
                        Freiburg,  Germany \\
%                     email: \url{wolfgang@kis.uni-freiburg.de} \\
              $^{5}$ Lockheed Martin Solar and Astrophysics Laboratory, Bldg. 252,
                        3251 Hanover Street,
                        Palo Alto, CA 94304, USA \\
%                     email: \url{e.mail-c} \\
              $^{6}$ Ingenieurb\"{u}ro f\"{u}r Optikentwicklung, Amalienstra\ss e 12, D-85737 Ismaning,
              Germany \\
%                    email: \url{e.mail-c} \\
                     }

\begin{abstract}
We describe the design of the \Sunrise Filter Imager (SuFI)  and the Image Stabilization
and Light Distribution (ISLiD)
unit onboard the
\Sunrise balloon borne solar observatory.
This contribution provides the necessary information which
is relevant to understand the instruments working principles, the relevant technical
data, and the necessary information about calibration issues
directly related to the science data.
\end{abstract}
\end{opening}

\section{Introduction}
\Sunrise is  a balloon-borne
solar observatory with a telescope with an aperture of 1~m, working
in the visible and near ultraviolet spectral domain. The main scientific goal of \Sunrise is to understand the
structure and dynamics of the magnetic field in the atmosphere of the
Sun.

The \Sunrise focal-plane science instrumentation consists of a Fabry-P\'erot filter magnetograph, and a
phase-diversity assisted filter imager working in the near UV, which takes
particular advantage from the observing conditions in the stratosphere.

\Sunrise provides high resolution ultraviolet images of the
photosphere and chromosphere, as well as Doppler- and magnetograms with an unpredecented
spatial resolution down to 0.15~arcsec on the solar surface.

\Sunrise is a joint project of the Max-Planck-Institut f\"ur
Sonnensystemfor-schung (MPS), Katlenburg-Lindau, with the
Kiepenheuer-Institut f\"ur Sonnenphysik (KIS), Freiburg, Germany,
the High-Altitude Observatory (HAO), Boulder, USA, the Lockheed-Martin Solar
and Astrophysics Lab. (LMSAL), Palo Alto, USA, and the spanish IMaX
consortium consisting of the Instituto de Astrof\'isica de Canarias (IAC),
the Instituto de Astrof\'isica de Andalucia (IAA), the Instituto Nacional de T\'ecnica
Aeroespacial (INTA), and the Grupo de Astronom\'ia y Ciencias del Espacio
(GACE).

The first stratospheric long-duration balloon flight of \Sunrise took place in
Summer 2009 from the Swedish Esrange station. For an overview on the \Sunrise project, instruments, and mission
we refer to Barthol  {\it et al.} (2010). In the present paper we describe one of
the two science instruments, the \Sunrise Filter Imager (SuFI) and the Image
Stabilization and Light Distribution (ISLiD) system, which were designed as a
single, integral unit.

\subsection{Science Requirements and Design Drivers}

The science goals of \Sunrise are described in Barthol  {\it et al.}  (2010) and are
not repeated here. Driven by these science goals,
the \Sunrise instruments are required to provide images of the magnetic structure and
measurements of the magnetic field, the flow velocity, and thermodynamic
properties of the plasma

\begin{itemize}
\item with a spatial resolution of order 0.1 arcsec,
\item on a field of view of 50$\times$50 arcsec (in the magnetograms),
\item over a sufficiently long time to follow the evolution of
magnetically active regions ({\it i.e.}, several days), and
\item simultaneously in different heights of the solar atmosphere.
\end{itemize}
This led to the concept of a
telescope of 1~m aperture operating in the visible and UV spectral ranges (down to
$\simeq210\,$nm), with in-flight alignment capability, equipped with a filter imager
and an imaging magnetograph, on a long-duration
stratospheric balloon flight in the framework of NASA's LDB program.

\subsection{\Sunrise Post Focus Instrumentation}

The \Sunrise post focus instrumentation consists of four units, two of which are science instruments,
the other two are system units for image stabilization and light
distribution.

\begin{itemize}
\item[a)]ISLiD

The \Sunrise science requirements demand simultaneous observations of the post focus instruments.
This is provided by ISLiD, the Image Stabilization and Light Distribution system of \Sunrise,
which will be described in detail in Section 2.
ISLiD contains a fast tip-tilt mirror, which is controlled by a correlating wavefront sensor
(CWS, Berkefeld {\it et al.}, 2010).
ISLiD is based on dichroic beam-splitters, which guide the different wavelength bands to the individual
instruments in the most efficient way. Part of the light, which is not used for scientific analysis is fed to the CWS.
In this way, simultaneous observations are possible with maximum photon flux in each channel.

\item[b)]SuFI

The \Sunrise Filter Imager (SuFI) samples the photosphere and
chromosphere in distinct wavelength bands. The channel at $214\,$nm allows
studies of the upper photosphere and lower chromosphere at a theoretical
angular
resolution of $0.05\,$arcsec (corresponding to $35\,$km on the Sun). Solar radiation in this wavelength
range is important for the stratospheric ozone household. The
OH-band at $313\,$nm and the CN-band at $388\,$nm provide high intensity contrast,
and thus sensitivity to the thermal structure of the photosphere and its embedded magnetic field structure.
The Ca {\sc ii} H line (singly ionized Calcium) at 397.6~nm is an excellent thermometer for the
chromospheric temperature structure.

\item[c)]IMaX

The Imaging Magnetograph eXperiment for \Sunrise (IMaX, Mart\'inez Pillet {\it et al.}, 2010) is an imaging
vector magnetograph based upon a tunable narrow-band Fabry-P\'erot filter. The instrument
provides fast-cadence two-dimensional maps of the  magnetic
vector, the line-of-sight velocity, and continuum frames with high
spatial resolution.
IMaX takes polarized images in two to five narrow wavelength bands in
either wing of the photospheric spectral line of Fe {\sc i} (neutral
iron) at 525.02~nm.

\item[d)]CWS

The Correlation tracker and Wavefront Sensor (CWS, Berkefeld {\it et al.}, 2010) is used for two purposes,
namely for precision image stabilization and guiding, and to control proper alignment of the telescope.

\end{itemize}

\section{Design of the SuFI/ISLiD Instrument}

\subsection{Optical Design Drivers and Design Philosophy}

ISLiD is a  complex optical instrument  that has to simultaneously
fulfill very different
tasks. Firstly, it has to stabilize the incoming beam from the telescope to provide stable output
images for the science instruments and the CWS. Secondly, it must divide the
incoming light into the wavelength bands required by the instruments, and
thirdly, it must feed the instruments with the corresponding wavelengths by
providing optical interfaces at given positions and with given directions,
plate scale, and pupil position. Since \Sunrise is a high resolution observatory,
two requirements must be satisfied: Firstly, the optical system must allow for
diffraction-limited performance under all environmental conditions that might arise during
flight. Secondly, the image stabilization system must damp the residual
pointing error (RPE) to a value which does not degrade the high intrinsic
image sharpness by image smear.

It is evident that the different instruments that are fed by ISLiD have
individual requirements, which makes it necessary to sort the instruments in a
priority list - not in the sense of ranking their scientific importance for the mission,
but rather from the point of view of their intrinsic challenges which they
impose on the optical design of ISLiD.

\begin{figure*}[t]
 \centerline{\includegraphics[width=1.\textwidth,clip=]{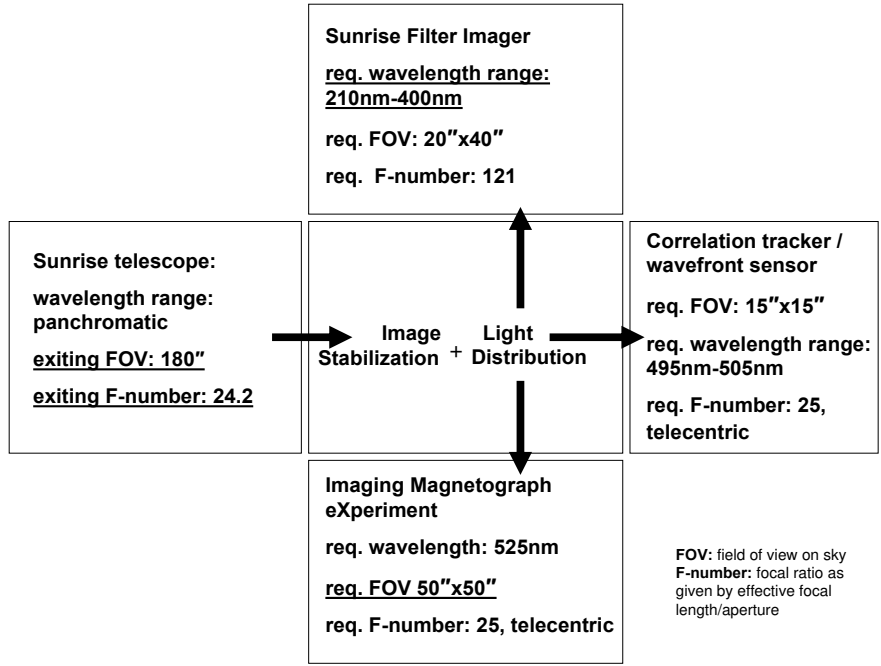}}
\caption{Scheme of the \Sunrise optical system with the relevant optical key
requirements form the point of view of ISLiD. Requirements driving the ISLiD design are underlined. See text for details.}
\label{islidscheme}
 \end{figure*}

 Figure~\ref{islidscheme} shows schematically the \Sunrise optical system and lists the key
optical requirements of the individual units from the ISLiD point of view.
The most severe requirement of each category is highlighted: the largest simultaneous field of view,
the broadest wavelength band, the shortest absolute wavelength, and the fastest working aperture given as effective F-number. Not listed
here are any further requirements such as beam direction, image position, or pupil
position.
Since all instruments afer designed to operate at their individual diffraction limit, which is
set by the telescope aperture and the specific working wavelength of the
instrument, the most critical instrument path is imposed by the SuFI
instrument, which aims at a diffraction limit at 214~nm. Also the wavelength
bandwidth of SuFI is by far the broadest, spanning almost a factor of two.
The SuFI science focus needs a five times larger plate scale than the telescope
exit focus (in order to allow for three-pixel oversampling of the point spread function
at 214~nm), and therefore a magnification of five is needed from telescope F2 to
SuFI science focus.
It is evident that diffraction-limited performance at these short wavelengths
can only be achieved if the number of optical surfaces is kept to a minimum.

To this end SuFI is completely incorporated into ISLiD, making the
SuFI image path an intrinsic part for all instruments. All instrument feeds make use of
the SuFI path, which is designed to be diffraction limited at 214~nm, and is therefore
uncritical for the longer wavelengths. The separation  of the UV light is done after the
beam stabilization, and close to SuFI focus, mitigating the disturbing
influence of the
 beam splitter plates, which must be expected to be bent out of shape by their
 complex coatings.
 After the separation of the UV light, each of the remaining instruments is fed by
 a dedicated optical feed providing the required magnification, image, and pupil
 position. Since CWS and IMaX are operating in the visible range and can be considered as monochromatic, this can
 best be achieved by dedicated lens optics.

 In order to achieve diffraction limited performance of the SuFI path,
 the following strategy was employed:
First, the number of
 surfaces needed to reimage (magnify) telescope F2 onto SuFI science focus must be minimized.
Further, the working field-of-view (FOV) of SuFI/ISLiD must be determined by the largest
 instrument field ($50'' \times 50''$) only and not by the very large telescope exit field of 180~arcsec, which is
 only required
 for image motion compensation.

This can be achieved by stabilizing the image {\it before} the light enters
the re-imaging optics: A field lens near the telescope focus
images the telescope aperture onto a plane folding mirror, which is actuated
by a fast piezo system and  acts as fast tip-tilt mirror. Behind this
mirror, we only have to consider the {\it reduced}  FOV of 50~arcsec in the design of the
imaging optics.
 With that, the remaining key requirements for the reimager are: A FOV of 50~arcsec, a
 working F-number of 24.2 in the incoming path, and an effective F-number of 121 in the
 exit path.
This allows for a design with two spherical mirrors only, which is optimum in
terms of design stability: The spherical mirrors are well polishable and
not very sensitive to alignment errors,  a fact that is of particular importance for
a balloon mission, where both, gravity load and thermoelastic deformations change with telescope elevation.

\begin{figure*}[t]
 \centerline{\includegraphics[width=1.\textwidth,clip=]{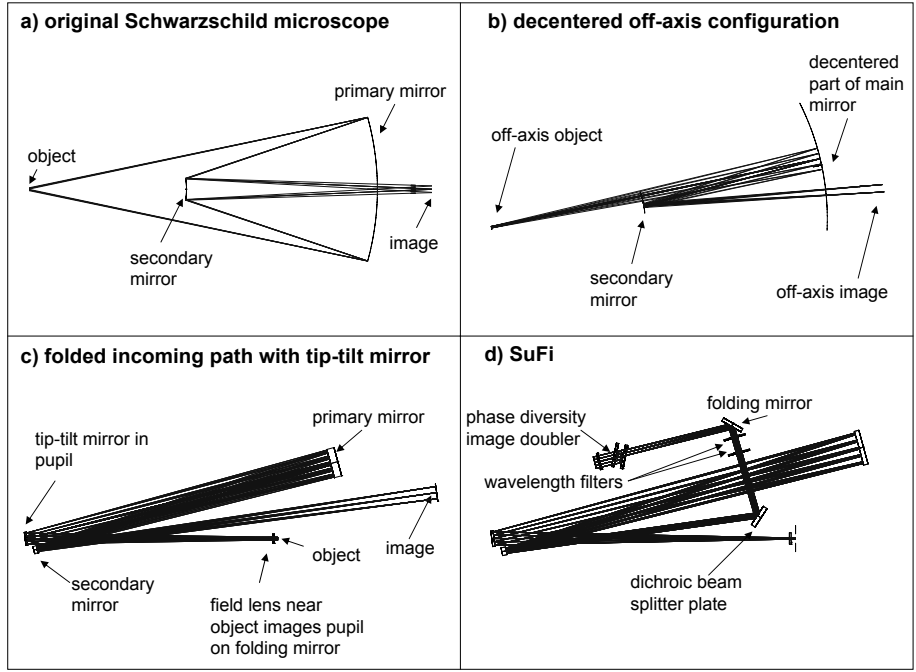}}
\caption{Optical design of SuFI: a) Principle of a Schwarzschild microscope.
The magnification of the sketched example design is 5, as in the SuFI instrument.
b) Sketch of the off-axis decentered pupil configuration.
The system is stopped down to an entrance F-number of 24.2, yielding an exiting F/121 beam.
c) Folded configuration of the Schwarzschild system.
A Lithosil field lens near the object plane images the telescope pupil on the folding mirror,
which can now be used as a fast tip-tilt mirror.
It compensates image motion in the object plane providing a fixed scene in image plane.
d) Final optical design of SuFI. The exiting beam of the Schwarzschild
microscope is folded by a dichroic beam-splitter plate, which reflects all wavelengths below 450~nm.
The folded beam passes a double filter set for wavelength selection before the beam is folded again
by a UV enhanced aluminum mirror. The phase-diversity image doubler is placed in front of the science focal plane.
It provides a focussed and an out-of focus image of the object side-by-side on the CCD detector.}
\label{steps}
\end{figure*}

\subsection{Optical Design}

We have chosen the design of a modified Schwarzschild microscope. The principle is sketched
in Figure~\ref{steps}a. Schwarzschild systems are common in ultraviolet microscopy. They
can have large numerical apertures and cover quite a large FOV. For a slow optical system with an F-number of 24.2
only,  the used aperture can be decentered and an off-axis
configuration can be used, thus eliminating any obscuration while preserving the superb
optical
performance.

The design steps from the principle of a Schwarzschild microscope to the
final SuFI arrangement are depicted in Figure~\ref{steps}.
From the original Schwarzschild configuration (Figure~\ref{steps}a) only a small part of the mirror
is used, corresponding to an entrance F-number of 24.2. This part of the mirror is
decentered (Figure~\ref{steps}b). In addition the system is used off-axis (object and image are not
on the optical axis given by the connection line of the mirror centers).
Figure~\ref{steps}c:
The incoming beam is folded by a small plane mirror, which is used
as image motion compensator (tip-tilt mirror). It must therefore sit in a
pupil of the system. This is achieved by placing a field lens near the telescope
secondary focus.
Figure~\ref{steps}d: A dichroic beam splitter plate separates the wavelengths below 450~nm by
reflection, while transmitting the longer wavelengths.
A set of interference filters in a filter wheel selects the working wavelength of SuFI.
Another folding mirror sends the light to the CCD.
In front of the CCD the phase diversity image doubler (see Section~\ref{pd}) is placed.

All components except the two spherical mirrors and the tip-tilt mirror are
close to focus and therefore optically not very critical. This is of key
importance especially for non perfect optical surfaces such as the dichroic beam
splitter and the interference filters (which are usually bent by the
coating),  as well as the 45 degree folding mirror. For these elements, the
surface quality requirements can be somewhat relaxed without compromising the
overall performance of the system.

\subsection{Phase Diversity Capability}
\label{pd}

\begin{figure}
 \centerline{\includegraphics[width=0.5\textwidth,clip=]{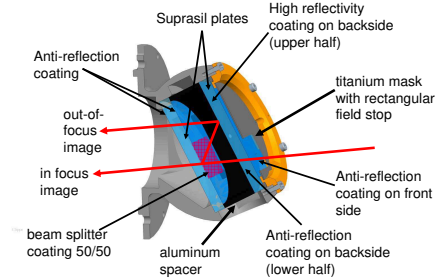}}
\caption{Design of the phase diversity image doubler. A field stop in front of the two glass plates selects the
useable field. Half of the light is transmitted through both plates directly and forms the in-focus image,
while the other half is optically delayed by two reflections at the specially coated inner surfaces.
The path difference is 28.15~mm, corresponding to an additional wavefront curvature of approx. half a wave at 388~nm and approx. one wave at 214~nm.
The two images are laterally separated by 12.5~mm, half the size of the CCD.   }
\label{dg}
 \end{figure}

Phase diversity is a technique for wavefront measuring and subsequent image
restoration (Paxman {\it et al.}, 1992; L\"{o}fdahl and Scharmer, 1994). Two images of the same
object, which are taken simultaneously at two distinct focal positions (usually one image in best
focus, the second defocussed such as to create an additional wavefront
curvature of typically one wave) allow the wave-front in the exit pupil to be sampled at two distinct focal positions.
From these two measurements, the object and the aberrations of the system can
be retrieved. While for static objects and aberrations the two images can be taken sequentially, for a
dynamic object like the Sun the two images must be taken simultaneously.
Phase diversity capability was incorporated in \Sunrise to correct for
 residual aberrations  due to thermoelastic
deformations of the telescope during flight.
To this end, the light path includes a  phase diversity image
doubler (see Figure~\ref{dg}). It resides directly in front of the focal plane and consists of two
plane-parallel Suprasil plates, which are coated with dedicated
high-reflectivity and anti-reflective coatings on both sides. The second
plate is coated on one half of its area with a 50/50 beam-splitter coating, that
transmits half of the incoming radiation, while reflecting the other
half. This leads to two images lying side-by-side on the focal plane: one image that has passed both
glass plates without any reflection, and another image, which has undergone two additional reflections, the first at
the beam-splitter plate, the second at the backside of the first plate.
The distance and the angle of the plates are chosen such that the path
difference between the images is 28.15~mm, while the lateral separation is
12.5~mm, corresponding to half the CCD size.
For the F/121 beam this path difference creates a defocus of approx. one wave
at 214~nm and half a wave at 388~nm.

\subsection{IMaX and CWS Feed Path}
Since the instruments have no internal focussing capabilities, ISLiD must ensure that
the exit foci for SuFI, CWS, and IMaX are all kept well in focus under all
environmental circumstances. Focussing of the entire postfocus assembly is
done by focussing the telescope in F2 by moving the telescope secondary
mirror (Barthol {\it et al.}, 2010).
The error signal for the focus position is provided by the CWS instrument (Berkefeld {\it et al.}, 2010). It is therefore
indispensible that the CWS and IMaX feed paths are identically affected by
thermoelastic distortions of the postfocus instrument platform and by
thermo-optic effects within ISLiD.
This is achieved by bringing the CWS optical interface into
close vicinity of the IMaX interface, and by using identical
optical components for both paths, which are separated by a beam splitter
plate. The coating of this plate reflects a band centered at 525~nm to IMaX
while transmitting a band centered around 500~nm to
CWS.

\begin{figure*}[b]
 \centerline{\includegraphics[width=1.\textwidth,clip=]{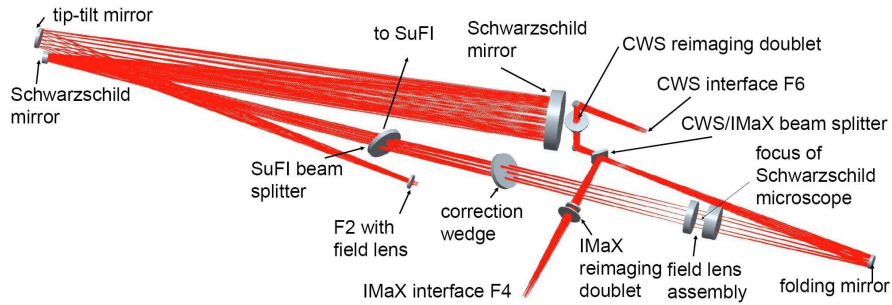}}
\caption{Optical design of the ISLiD-IMaX and ISLiD-CWS feed paths. The exiting beam of the Schwarzschild
microscope is transmitted by a dichroic beam-splitter plate, which reflects all wavelengths below 450~nm to SuFI.
A wedged corrector plate is used to compensate for astigmatism and coma induced by the SuFI beam splitter.
In the Schwarzschild focus a field lens assembly is placed, creating a real pupil image. In
the folded path a second lens doublet reimages the Schwarzschild focus
at the required position, while sending the pupil to infinity to fulfill the requirement of telecentricity. }
\label{islidfull}
 \end{figure*}

\begin{figure*}[t]
 \centerline{\includegraphics[width=1.\textwidth,clip=]{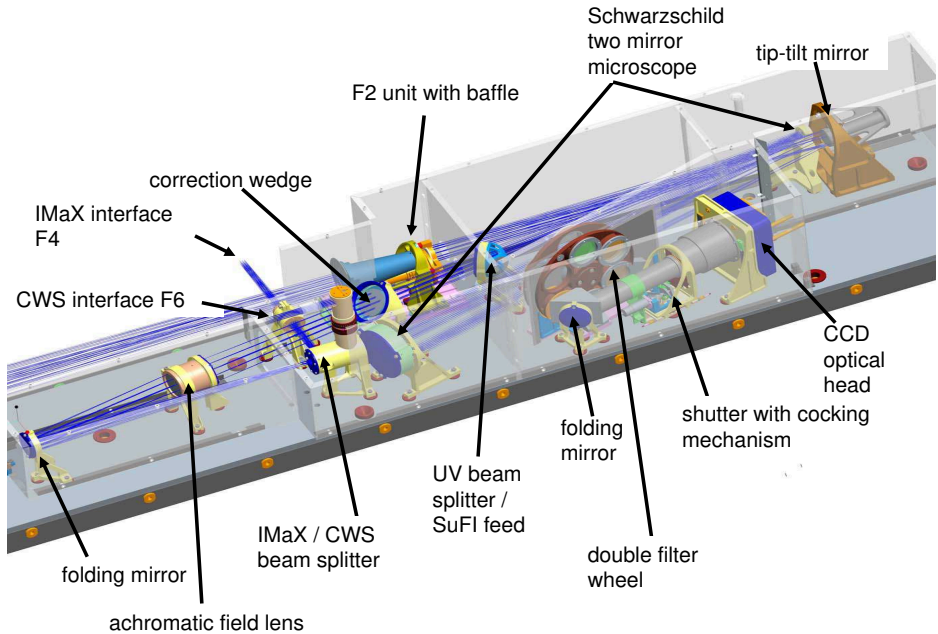}}
\caption{CAD overview of the ISLiD/SuFI optics unit,  showing the SuFI path.
Housing is shaded to provide insight into the compartments.}
\label{Cadleft}
 \end{figure*}

\begin{figure*}[]
 \centerline{\includegraphics[width=1.\textwidth,clip=]{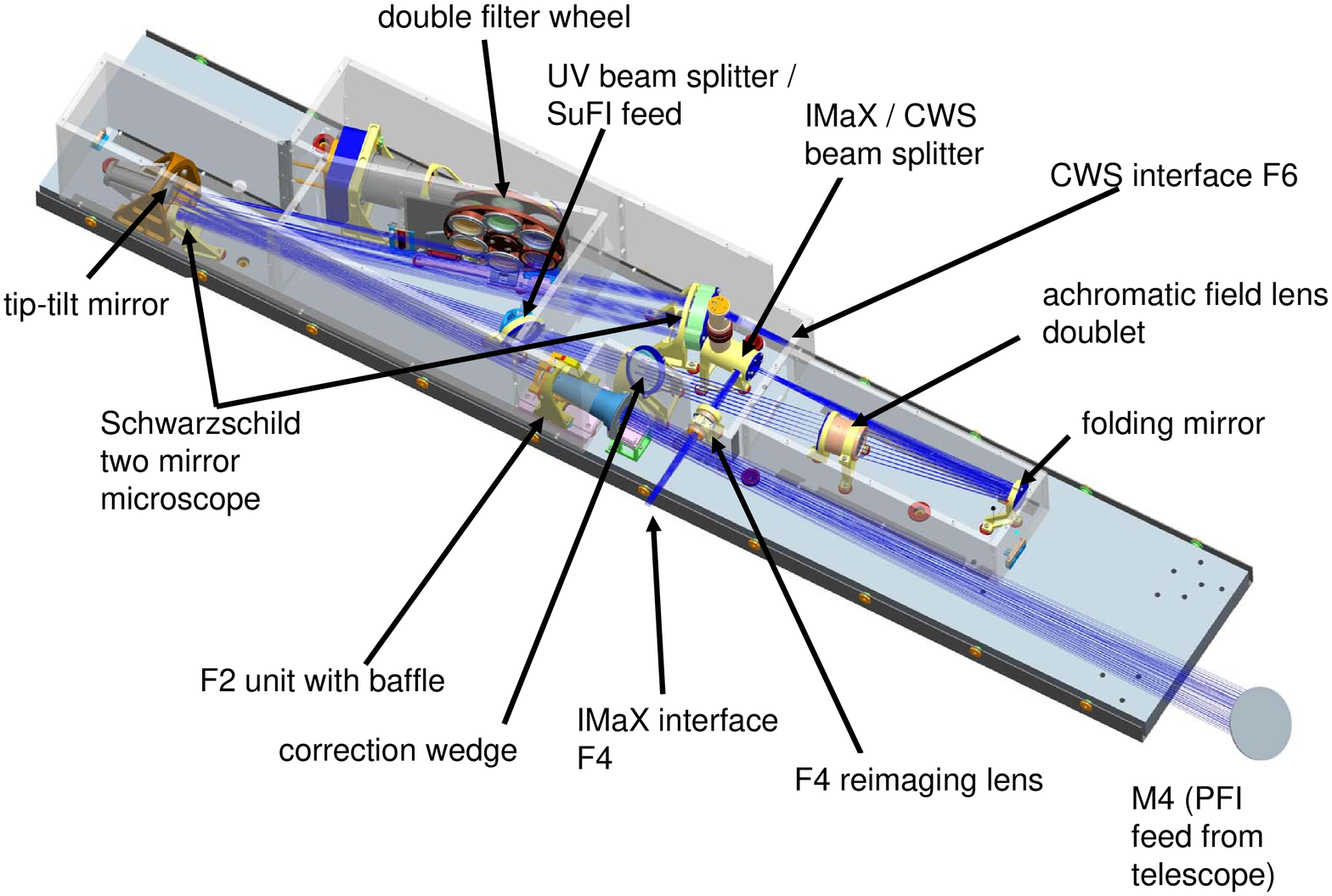}}
\caption{CAD overview of the ISLiD/SuFI optics unit.  Light from the telescope is fed by
folding mirror M4 into the system. Telescope secondary focus is at F2, where the field is
stopped down to 180 arcsecs. The  optical path of ISLiD is seen with the  foci F4 and F6 feeding
IMaX and CWS, respectively.}
\label{Cadright}
 \end{figure*}

Both paths start from a common achromatic field lens assembly in the focus
of the Schwarzschild microscope, which has been corrected for astigmatism and coma induced
by the SuFI beam splitter plate by inserting a wedged corrector plate (see Figure~\ref{islidfull}). The field lens forms a real image of the
instrument pupil. A second lens assembly reimages the Schwarzschild focus
with the required plate scale (demagnification by a factor of 4.85), while creating a pupil image at infinity, as
required by the instruments (telecentric exit paths).
The full ISLiD optical design showing the ISLID-IMaX-CWS feed paths is depicted in
Figure~\ref{islidfull}. Figure \ref{Cadleft} shows the mechanical implementation
of SuFI/ISLiD looking onto the SuFI compartment, while Figure~\ref{Cadright}
depicts the view from the other side, allowing a view onto the CWS and IMaX
feed path implementation.

Besides the aspects discussed so far, the IMaX feed path must conserve
the high polarimetric efficiency of the IMaX instrument. This implies that ISLiD
must not produce significant linear polarization. Since in the design of ISLiD the
possibility to feed an additional instrument in future is foreseen (a near IR
full Stokes spectropolarimeter with a working wavelength at 854.2~nm), ISLiD
was designed for minimum linear polarization at 525~nm and 854~nm. This was
achieved by using metallic surfaces  close to normal incidence and by a careful
design of the beamsplitter coatings.

\subsection{ISLiD Mechanisms: F2 Unit and Tip-Tilt Stage}

ISLiD contains two mechanisms related to the CWS unit:
The tip-tilt mirror stage and a calibration filter wheel in F2.
The tip-tilt stage is a fast piezo-stage (Physik Instrumente PI S330K032)
with  a range of $\pm$~50 arcsec on the sky. It is described in detail in
Berkefeld {\it et al.} (2010).

\begin{figure}[h]
 \centerline{\includegraphics[width=0.5\textwidth,clip=]{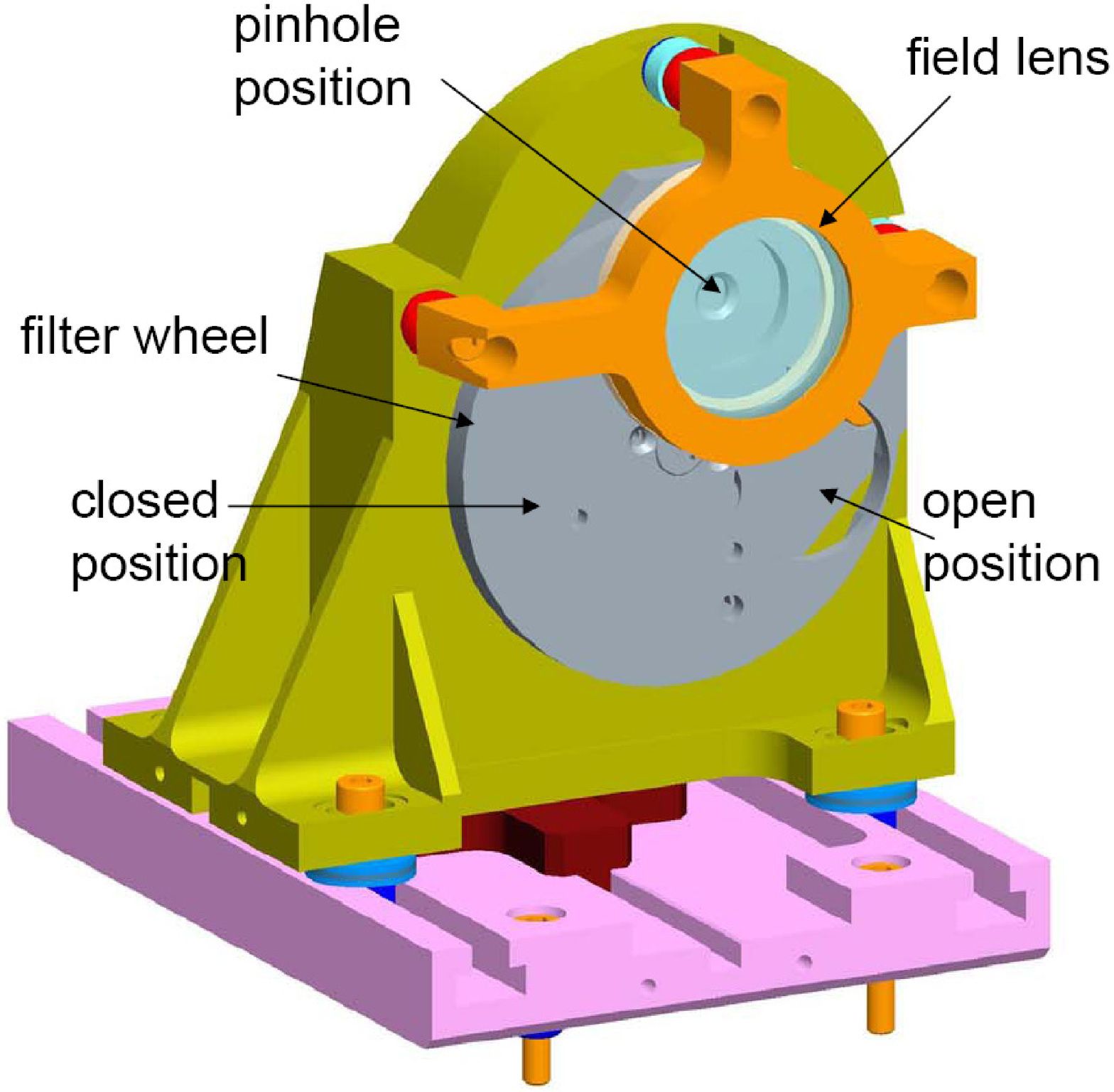}}
\caption{CAD design of the F2 calibration unit. The field lens is  mounted onto the same mount.}
\label{f2}
 \end{figure}

The F2 calibration unit (see Figure~\ref{f2}) is a motorized filter wheel with three positions: An
open position, a closed position for dark exposures, and a pinhole, which is
used to monitor the relative alignment of the science instrument fields, and
which is needed for CWS internal calibration during flight.

\subsection{SuFI Mechanisms: Filter Wheel and Shutter}
\subsubsection{Wavelength Selection}

The spectral bands to be observed by SuFI are summarized in Table 1.

\begin{table}
\begin{center}
\begin{tabular}{lll}
\hline
{\bf 214~nm continuum} & center wavelength & 214~nm  \\
             & FWHM & 10~nm  \\
            & number of filters & 2  \\
            & exp. time   &  30~s   \\ \hline
 {\bf 300~nm continuum} & center wavelength & 300~nm  \\
             & FWHM & 5~nm  \\
            & number of filters & 2  \\
            & exp. time   &  300~ms   \\  \hline
 {\bf 312nm continuum } & center wavelength & 312~nm \\
 {\bf with lines of OH}            & FWHM & 1.2~nm  \\
            & number of filters & 2  \\
            & exp. time   &  200~ms   \\     \hline
 {\bf 388~nm CN band head} & center wavelength & 388~nm \\
             & FWHM & 0.8~nm  \\
            & number of filters & 2  \\
            & exp. time   &  100~ms   \\      \hline
 {\bf 397nm Ca {\sc ii} H} & center wavelength & 396.8~nm \\
             & FWHM & 0.18~nm  \\
            & number of filters & 1  \\
            & exp. time   &  1~s   \\

\hline
\end{tabular}
\caption{Filters and typical exposure times used in the SuFI instrument}
\end{center}
\end{table}

The sunlight intensity decreases strongly towards shorter wavelengths so
that in the case of the UV wavelengths the suppression of unwanted
contributions from leakage at longer wavelengths is only achievable by combining
two filters at each filter position. A single specially blocked filter is sufficient only for the Ca {\sc ii} H
line.

The filters are combinations of dielectric interference pass bands with color glass
blocking, custom made by Barr Associates. All dielectric coatings have been applied using
ion-assisted
deposition (IAD) techniques, which greatly helps in reducing the temperature
sensitivity of the bandpass down to 0.005~nm/$^\circ$C. This is of particular importance in a balloon
mission, for which quite a large temperature range must be assumed during flight.

The tiny optical thickness variations of the different filter sets are well within
the depth of focus of the F/121 beam.

The filters are placed approx. 25~cm in front of the focal plane, which for
the F/121 beam is sufficient to prevent pinholes and other effects of the
filters to be imaged onto the CCD, while at the same time being close enough
to
make the image insensitive to wavefront deformations induced by the filters.

In order to minimize the induction of any torque into the instrument
the two filters are mounted in two parallel separate filter wheels, which are driven
by a common drive system in
opposite directions.
Each of the  filter wheels has six positions. The
filters are mounted with an inclination of 0.5
degrees to avoid ghost images. Between the two filter wheels, a baffle is
placed, which is part of the inner baffling/housing system (see Section~\ref{baff})
that optically seals the optical path after wavelength selection.

The six positions of the filter wheels are used as follows: The four filter pairs (214~nm,
300~nm, 312~nm, and 388~nm) are  arranged in such a way that the
corresponding filters of the two wheels are always in the light beam at
the same time. The single Ca {\sc ii} H filter (397~nm) is placed in the
"sunny side" filter wheel (for stray light reasons) whereas the
corresponding camera side filter wheel position is left empty. The
remaining filter position contains a  combination
of  lenses and neutral density filters,
which image the tip-tilt mirror and thus the pupil onto the CCD. This feature was mainly needed
during the PFI integration on top of the telescope, but
provided also a monitor of the pupil
(which is sensitive to the relative alignment of ISLiD with respect to the telescope) during flight.

\Sunrise science requires series of alternating images of the solar
photosphere and chromosphere with a high cadence.
This means that changes between 397 and 300~nm, between 397 and 388~nm,
and between 214 and 313~nm must be done as fast as possible. This can
only be achieved if the respective filter combinations are placed at
neighboring filter wheel positions.

The filter wheel is operated  such that after one wavelength cycle
the whole wheel is brought back to the first filter position not by
continuation in the same rotation direction (+60 degrees), but by backward rotation (-300 degrees). This
ensures a higher positioning accuracy even after many thousand cycles as
expected during a several day mission.

\subsubsection{Mechanical Shutter and Exposure Control}

\begin{figure}
 \centerline{\includegraphics[width=0.5\textwidth,clip=]{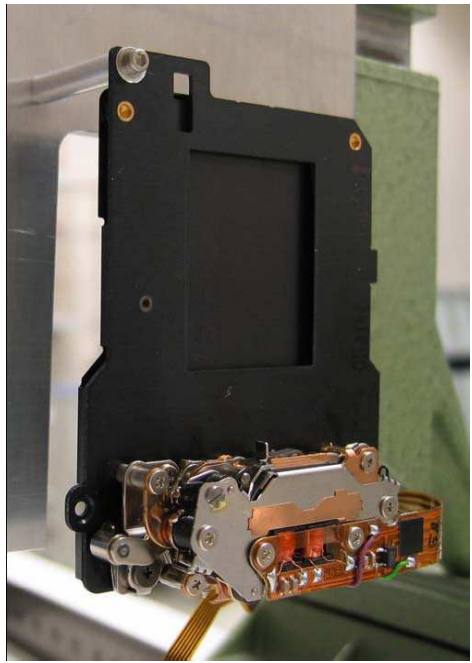}}
\caption{Nikon D2H focal plane shutter as used in the SuFI instrument.}
\label{shutter}
 \end{figure}

Since the full area of the SuFI CCD is used for exposure, and thus the CCD
cannot be operated in frame transfer mode,
a separate mechanical shutter is used
for exposure control.
This shutter is placed near the focal plane in front of the phase diversity
image doubler. After different commercially available shutters had been
tested, a Nikon dual-blade focal plane shutter as used in the Nikon D2H type cameras
was selected (see Figure~\ref{shutter}). It allows for fast exposures,
exposes the covered area evenly and is easy to control. However it needs
a mechanism for cocking.

\begin{figure}
 \centerline{\includegraphics[width=0.5\textwidth,clip=]{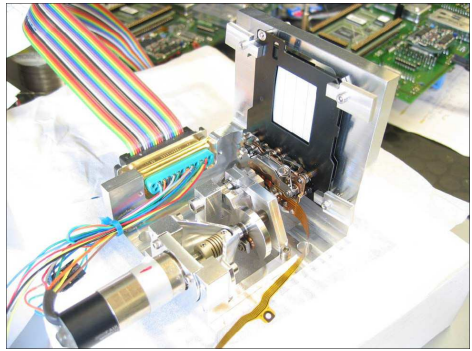}}
\caption{Cock mechanism for the SuFI shutter.}
\label{cock}
 \end{figure}

The cock mechanism is shown in Figure~\ref{cock}.
A motor with an attached 76:1 gear drives a spiral, operating a rocker. The rocker has a wheel
where it is operated by the spiral. On the other end it operates the cocking
lever of the Nikon shutter. There is a small
Vespel part between the rocker and the shutter cock lever to reduce the point
pressure and wear of the metal surfaces.

In order to mitigate disturbing effects by shutter cocking and release on the
instrument, the whole shutter assembly is mounted on spring-loaded steel
wires, which significantly damps the mechanical shock load on the ISLiD base plate when the shutter blades
are released.

\subsection{SuFI CCD Sensor and Camera}
The camera of SuFI must fulfill the following key requirements: it must be
UV sensitive down to 210~nm, and it must combine a large format of 2048
$\times$ 2048 pixels with a fast read-out speed, such that at least one full frame
per second can be read.
We have chosen a PixelVision BioXight BV20CCD camera, based on a
   SITe S100AB-04 CCD with
            2048 $\times$ 2048 pixels and
           12~$\mu$m x 12~$\mu$m pixel pitch. The sensor is
            backside illuminated.
            It is used in full frame mode, with four output ports.

The camera electronics provide     14-bit digitization
             at a frame rate of a maximum of 2.64 frames per second.
            The commercial version of this camera employs a thermoelectric cooler with a Peltier element and
             with liquid cooling of the Peltier warm side.
            Specified dark current is 1.82~electrons per pixel per second at a CCD temperature
            of 240 K.

For utilisation in \Sunrise the camera was heavily modified: The  camera
head (CCD) and camera electronics were separated in order to minimize
heating of the camera head by power dissipation in the camera electronics.
This separation allows two independent temperatures and cooling systems for the two subsystems.

While the camera head
is cooled to about 270~K by two custom made methanol heat pipes
(by Advanced Cooling Technologies (ACT), USA) directly connecting the warm side of
the Peltier element with a distinct radiator, the camera electronics are
housed in a pressure vessel,  one side of which is used directly as a radiator. In this
way the camera electronics (and thus the radiator) can operate at a significantly higher temperature
than the CCD head, resulting in a high cooling efficiency of the radiator.

Electric connection between CCD and camera electronics is established by
vacuum feedthroughs.

\subsection{Straylight Control}
\label{baff}
The solar surface represents an object with relatively low intrinsic intensity contrast.
The observed contrast values are further diminished by optical aberrations
({\it e.g.} Danilovic {\it et al.}, 2008) and instrument stray light ({\it e.g.} Mathew {\it et al.},
2009). Both factors are of increasing importance when observing in the
UV for different reasons: optical aberrations due to optical surface
imperfections scale with  $\lambda^{-1}$. At the same time {\it spatial} scattering (angular
redistribution at optical surfaces) is typically also more pronounced at shorter
wavelengths. The most important source of straylight in solar UV imagers is, however,
{\it spectral} contamination by photons of the longer wavelength parts of the
photospheric
 spectrum, which can be orders of magnitude more intense than the photon
contributions within the selected UV science passband.
To suppress these contaminations  two identical filters in tandem are used, which has the additional
advantage of providing a more "rectangular" band pass shape.
Behind the filters the beam is highly sensitive to parasitic light and must
therefore be completely light-tight. This is ensured by a baffle tube between
filter wheel and camera head. Folding mirror, shutter assembly, and phase
diversity image doubler are all contained in this tube system.

\begin{figure*}
 \centerline{\includegraphics[width=1.\textwidth,clip=]{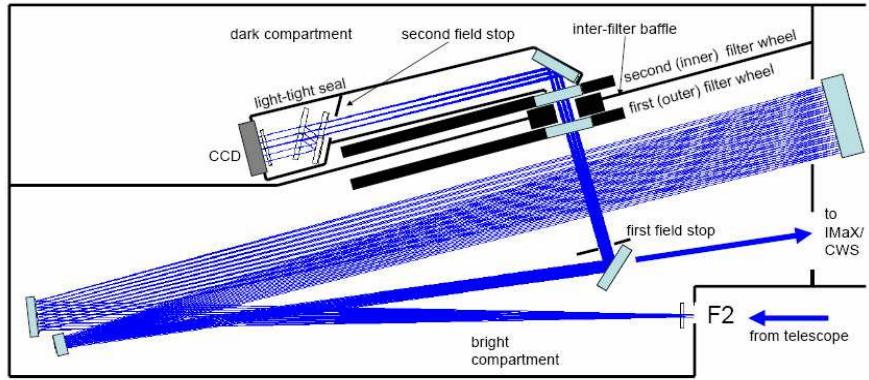}}
\caption{Scheme of the baffling system of SuFI. The light-tight innermost seal
covers the light path after the second filtering, which is most vulnerable to parasitic illumination.
The double filter wheel seperates the dark compartment from the bright compartment. Two consecutive field stops narrow the field
which is propagated to the detector.}
\label{baffle}
 \end{figure*}

In addition to this innermost seal, the whole ISLiD unit is divided into
several compartments, which are separated by baffles and walls. The baffling
concept is depicted in Figure~\ref{baffle}.
The first compartment is the brightest one. It serves as a light trap for the
unused portions of the telescope field-of-view of 180~arcsec, which illuminates the
tip-tilt mirror but is not completely picked up by the limited field of view of the Schwarzschild
system (60~arcsec). This is still significantly larger than the SuFI science field of view of $17'' \times 35''$. Therefore, another mask is placed in the reflected beam of
the UV separator, such that the filter wheel is illuminated by a slightly vignetted field of $20'' \times 40''$ only, with $17'' \times 35''$ unvignetted.
The filter wheel provides a separation between the bright compartment and a dark compartment, which is a closed
box containing the inner filter wheel and
the camera head, together with the above-mentioned light tight innermost baffling
tube. The first filter wheel resides in the bright compartment, and a baffle
tube between the two filter wheels forms the optical feed-through from one compartment
into the next.

\section{Structural and Mechanical Design of SuFI/ISLiD}
\subsection{Postfocus Instrumentation Platform Concept}

ISLiD/SuFI, CWS, and IMaX are built as individual optomechanical subunits (modules).
The general design approach of the Postfocus Instrumentation platform (PFI) is to accommodate, implement, and align these individual modules into the PFI
support structure. All modules are internally aligned and tested before integration into the PFI support
structure. The concept of individual modules has the advantage that each module can be designed to achieve maximum
internal stiffness, which helps to minimize internal alignment errors due to changing gravity load (elevation change).
The IMaX and CWS modules are isostatically mounted into the support structure. Forces arising from
differential thermal expansion between the modules and the PFI
thus do not lead to mechanical tensions in the modules.
Relative lateral alignment of the individual modules within the support structure must be guaranteed
within limits of $\pm$0.2~mm in all directions. This requirement arises from the maximum image shift that
can be tolerated at the optical interfaces between the modules.
Angular alignment of the modules with respect to the fixation point (boresight error) must be ensured
within a limit of 3~arcmin in all directions, compliant with a maximum relative pupil shift
of 5\% of the pupil diameter.
\subsubsection{ISLiD as Alignment Master Unit}
Since ISLiD has four optical interfaces, each described in position and in/out beam direction, it is
convenient to use ISLiD as the `alignment master': ISLiD is mounted rigidly within the
instrument platform PFI; the relative alignment of ISLiD with respect to the telescope is done in
the opposite sense: the telescope is aligned to ISLiD/PFI via the plane folding mirrors M3 and M4,
which help in adjusting the incoming beam position (lateral as well as focus) and beam direction
(pupil position). The ISLiD entrance F2 position is exclusively determined via the internal
alignment of ISLiD (internal foci and stops) and may not be regarded as absolute in position.
It is used as a focus compensator during alignment.

\subsubsection{PFI Mechanical Structure}
In order to maintain the optical alignment of the modules within the PFI
for different thermal and mechanical load cases during flight, the PFI is built
from carbon fiber reinforced plastic (CFRP) honeycomb plates, which are
connected via aluminum inserts and screws. Such a structure has a very high
stiffness-to-weight ratio.
\subsubsection{Thermal Aspects}
The CFRP plates were designed for minimum
thermal expansion. The ISLiD optical bench plate (made from the same
material) provides an additional stiffening of the PFI.
The CWS base plate is also made from the same material, mitigating relative
thermal expansion of the module within the PFI. Only the IMaX module is made
from aluminum honeycomb; however, the effects of relative thermal expansion
is uncritical, owing to the isostatic mounting of the IMaX module within the
PFI platform.
 The Schwarzschild
system itself is not perfectly athermal. It is based on Schott DURAN glass for the Schwarzschild
mirrors in combination with the low thermoelastic expansion coefficient of the CFRP baseplate.
The refractive reimaging paths feeding IMaX
and CWS, respectively, make use of aluminum spacers to compensate the
change in refractive index of the glass used for the lenses. The most
important aspect is, however, that the two feed paths of IMaX and CWS are
identically constructed and put to close vicinity, which minimizes the risk of a relative defocus between the
two modules even in case of temperature inhomogeneities within the  PFI.

The  cameras and electronics inside the PFI
 need to be radiatively cooled. The PFI modules are therefore equipped with
 radiators tailored to achieve acceptable operating temperatures for the camera head (between -5$^\circ$C and +10$^\circ$C) as
 well as for the electronic components (below 45$^\circ$C) nearby. The required radiator sizes and surface properties were derived
 from thermal analysis taking into account the flight scenario and the required operating
 temperatures.

\section{Instrument Operation, Observing Modes, and Science Data Products}

   \subsection{Observing Modes}

   The SuFI software allows for flexible observing programs. Observing modes
   differ basically in the wavelengths which are observed. For example, the
   shortest wavelength of 214~nm should only be observed when the elevation of
   the Sun is larger than 30~degrees above the horizon, and when the balloon
   altitude is above 36~km in order to minimize atmospheric extinction. By skipping the
   shortest wavelength the effective cadence of the other wavelength observations
   can be increased from 1 exposure per 40~s to 1 exposure per 8~s for a full wavelegth cycle.
   For highest cadence observations a fixed
   wavelength observing mode is available for each wavelength (1 exposure per second). Also an alternating
   wavelength mode between two or more filter positions is possible for each
   combination of wavelengths. Besides the used wavelengths with their specific
   exposure times, there are several other factors that limit the image cadence:
   the filter changing time (1~s), the cocking time of the mechanical shutter (0.5~s), the CCD
   readout time (0.4~s), the execution time of the image compression, the network
   bandwidth, and the performance of the SuFI and the Instrument Control Unit
   (ICU) software.

   \subsection{Software Design}

   The system design of SuFI is mainly driven by the requirements coming from
   the PixelVision camera, which, for example, constrains the operating system to
   Microsoft Windows 2000. Other subsystems of the \Sunrise mission are constrained
   to other operating systems, which leads to a heterogeneous software environment.
   In a nearly autonomous mission of six days duration, robustness and reliability
   of all system components, in particular of all software components, are
   mandatory. We therefore use the Adaptive Communication Environment
   (ACE) library, whose reliability is well proven by many applications running
   on several platforms over more than a decade (Schmidt and Huston, 2002,2003).
   The core of the pattern-oriented ACE library is a thread pool reactor
   (Schmidt {\it et al.}, 2000) that transfers scientific images as well as house-keeping
   data over the 100~MBit/s Ethernet network connection to the ICU, which processes
   the data and stores them on one of two data storage subsystems (DSS) via
   an IEEE 1394 connection.  Optionally, data are sent to the ground over one
   of two telemetry channels. Additionally, the reactor receives commands
   from  two possible command sources: a) Commands can be sent from a timeline-
   driven observing program stored on the ICU before launch and b) commands can
   be sent manually over a telemetry channel from the ground.

   Since multithreaded programming in a distributed system is difficult and
   error-prone, we developed a memory checking tool to detect memory leaks automatically
   before launch as well as a call stack logger tool to detect deadlocks. Our
   aim of a maximum performance of the SuFI software ({\it i.e.} the maximum image cadence)
   is reached by  a profiling tool that  measures the
   execution time of each critical part of the software. Our development of both,
   the call stack logger and the profiler, are based on the technique of method
   call interception (Sayfan, 2005).

   \subsection{SuFI Electronics Unit}

   The SuFI Electronics Unit (SuFI-EU) is an IBM compatible computer system with
   all the interfaces needed to connect the other hardware components of the
   SuFI instrument. The heart of the SuFI-EU box is an AMPRO LittleBoard 800
   mainboard, featuring an Intel 1.4~GHz LV Pentium M 738, 1~GB RAM, 4~GB flash
   disk boot device. A 100~MBit Ethernet port  connects the ICU, a PCI slot
   connects the PixelVision Lion~2 frame grabber card, which provides the
   interface to the CCD camera. A serial port RS422  connects the SuFI
   mechanism controller responsible for the filter wheel, the shutter, and the
   camera's exposure, and a serial port RS232 connects the internal Health
   Monitoring System for temperature and pressure monitoring and for control of
   the air circulation inside the hermetically sealed SuFI-EU box via three fans.

   \subsection{Science Data Products}

   During the six-day duration mission in June 2009 the SuFI instrument acquired about 150\,000 images.
   Each image is compressed and a header with a current set of house-keeping
   data is added. These raw data are transferred to the ICU that stores them
   in a RAID level~5 format on the DSS. After the successful landing, the DSS was
   recovered, the RAID format was decoded, and the raw image data format was
   converted into the level~0 format using the Flexible Image Transport System
   (FITS). The FITS header consists of two sections of entries. The first section
   is the common part of SuFI and IMaX and contains system information like
   acquisition time, reduction level, pointing position, solar ephemeris data,
   GPS position and flight altitude, etc. The second section contains the
   instrument specific part and is different for SuFI and IMaX.

   The level~0 science data are further processed by correcting for dark current
   and flat field (acquired in flight by averaging 100 exposures during a 10~min random walk on the solar surface using the coarse pointing system) and by removing some instrumental effects, which provides
   the
   level-1 FITS data. Finally, the phase diversity reconstruction leads to
   the level-2 FITS data.

\section{Optical On-Ground Performance}
\begin{figure}
 \centerline{\includegraphics[width=0.8\textwidth,clip=]{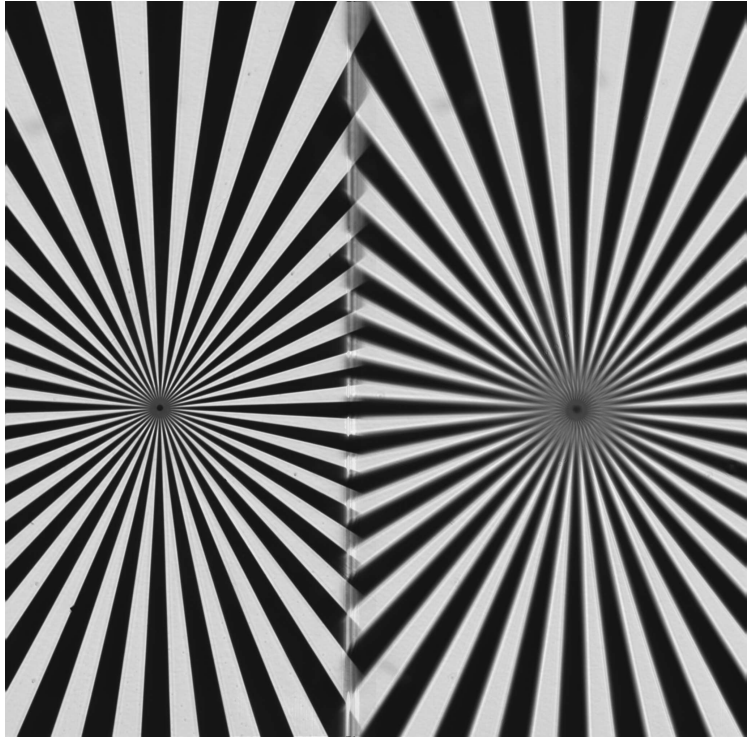}}
\caption{Image of a Siemens star target located in telescope F2 taken in the 388~nm channel.
The differential defocus between the sharp image (left) and the defocussed image (right) is $\lambda$/2.
The cut-off frequency is consistent with the theoretical diffraction limit. }%\label{fig:?}
 \end{figure}
\begin{figure}
 \centerline{\includegraphics[width=0.8\textwidth,clip=]{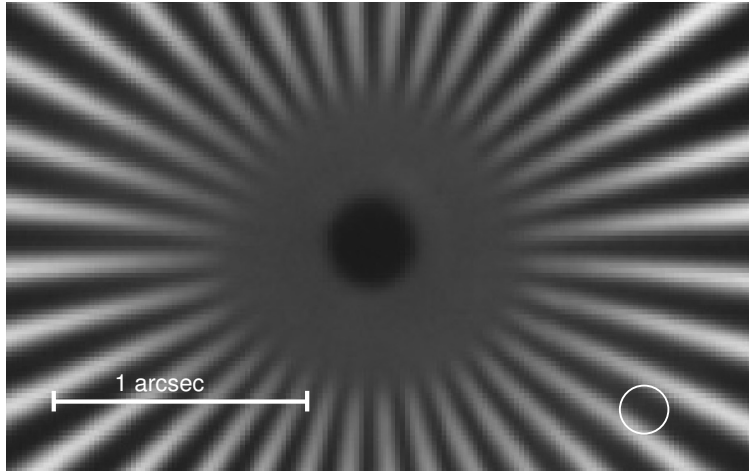}}
\caption{Detail from above, showing the innermost part of the in-focus image. The white circle indicates the classical Airy
disk as defined as 1.22$\lambda/D$, which for the test wavelength of 388~nm corresponds to a radius of 0.1''.
At this wavelength the diffraction limit is strongly oversampled, which can clearly be seen. }%\label{fig:?}
 \end{figure}

ISLiD/SuFI was tested and calibrated on ground prior to the science flight.
The optical performance of SuFI was verified by inserting optical
targets in F2 and analysing the images of the targets taken by the SuFI
camera. For the ISLiD-IMaX and ISLiD-CWS feed paths a dedicated camera placed in F4 and F6, respectively, was used.
Contrast transfer and spatial resolution (cut-off frequency) was
tested with a Siemens star target. Distortion was measured with the
help of a rectangular grid target. Internal magnification and anamorphotic distortion
were measured by a micrometer scale in F2.
The optical quality of the ISLiD-IMaX feed path was measured in terms of
wavefront distortion by use of an interferometer in autocollimation setup,
thus in double pass. At the measurement wavelength of 633~nm, the rms wavefront
error in double pass including the autocollimation sphere and the interferometer was $\lambda$/22.
Polarimetric properties of the ISLiD-IMaX feed path were characterized in
terms of Mueller matrix maps over the full IMaX field of view by inserting
 polarization optics (rotatable linear polarizer and quarter wave plate) in front of F2 and measuring the
 full Stokes vector
image at F4 with the Zurich imaging polarimeter ZIMPOL~{\sc ii} (Gandorfer {\it et al.}, 2004) at a
wavelength of 525~nm selected by an interference filter of 5~nm width.

Both, the polarimetric calibration and the interferometric test of the
wavefront distortion confirmed the predicted performance. As seen
from IMaX, ISLiD is basically invisible in terms of wavefront degradation as
well as degradation of the very high IMaX polarimetric efficiencies (Mart\'inez Pillet {\it et al.},
2010).
Some optical key performance values are listed in Table 2.

\begin{table*}
\centering
    \begin{tabular}{ lcccl}
    \hline
    Item & Wavelength & Design goal  & Measured & Comment  \\ \hline
       Internal  &  500~nm   &  1.05   &  1.041 &  measured with\\
       magnification &    &     &     &     micrometer scale \\
       &    &     &     &     in F2 and in F6 \\ \hline
         &  525~nm   &  1.05   &  1.034 &  measured with\\
         &   &    &     &     micrometer scale\\
         &  &     &     &      in F2 and in F4   \\ \hline
         Transmittance  &  214~nm   &  $\ge$0.5    &  0.52  & excluding filters; \\
         &  &     &     &     calculated from  \\
         &  &     &     &     transmission \\
         &  &     &     &      measurements of \\
         &  &     &     &     components / samples \\ \hline
           &  525~nm   &  $\ge$0.5    &  0.56  &  measured \\
           &      &     &     &     between F2-F4 \\ \hline
           &  500~nm   &  $\ge$0.5    &  0.56  &  measured \\
           &      &     &     &     between F2-F6 \\ \hline
        Linear    &  525~nm  &  $\le$0.1   &  0.04 & measured   \\
        polarization &  &     &     &   \\ \hline
         Distortion &  all   &  $\le$0.01   &   $\le$0.005 & measured with \\
         & wavelengths &      &    &     grid target in F2  \\ \hline
        WFE (RMS)  &  633~nm  &      &  $\lambda$/22 & measured in \\
        on ground &   &     &     &     double path at F2-F4;\\
        &   &     &     &      focus removed   \\ \hline
         WFE (RMS)  &  300~nm  &      &  $\lambda$/17 & deduced from \\
        in flight &   &     &     &     PD analysis\\
        &   &     &     &      telescope removed   \\ \hline
       WFE (RMS)  &  300~nm  &      &  $\lambda$/5-6 & deduced from \\
        in flight &   &     &     &     PD analysis\\
        &   &     &     &      telescope included   \\ \hline
         Relative    &  500~nm  &  $\le \lambda$/10     &  $\lambda$ /20 & in-flight \\
         defocus &   &     &     &     performance, \\
        SuFI-F6 &   &     &     &      from PD    \\ \hline
            Relative    &  525~nm  &  $\le \lambda$/10     &  $\lambda$ /20 & in-flight \\
         defocus &   &     &     &     performance, \\
       SuFI-F4 &   &     &     &      from PD    \\ \hline
\end{tabular}
\caption{Key optical performance parameters of ISLiD/SuFI.}
\label{table:2}
\end{table*}

\section{Conclusions}
The instrumental concept and design of ISLiD/SuFI was finally proven
during the first scientific balloon flight of the \Sunrise mission from 8
to 13 June 2008.

During the whole period the instrument performance was excellent. IMaX and SuFI
were coaligned and both in focus, which was actively controlled by CWS (by focussing the telescope).
Relative defocus between the different exit foci of ISLiD was always below
$\lambda$/20, well in the range that can be corrected for during phase diversity
treatment of the data.
IMaX wavefront performance was not affected by ISLiD, and the polarimetric
efficiency of IMaX was always nominal, i.e. the Mueller matrix of ISLiD was
stable (Mart\'inez Pillet {\it et al.}, 2010).

SuFI collected a total of 150\,000 images in all wavelength bands. Except for a
negligible
number of excursions the SuFI shutter worked perfectly; the filter wheel
had no functional problem at all.

SuFI collected the first high resolution (order 0.1~arcsec) images of the solar photosphere in
the important wavelength range between 200~nm and 350~nm, which is of particular interest
for solar irradiance modelling.

\begin{figure*}
 \centerline{\includegraphics[angle=90,width=1.\textwidth,clip=]{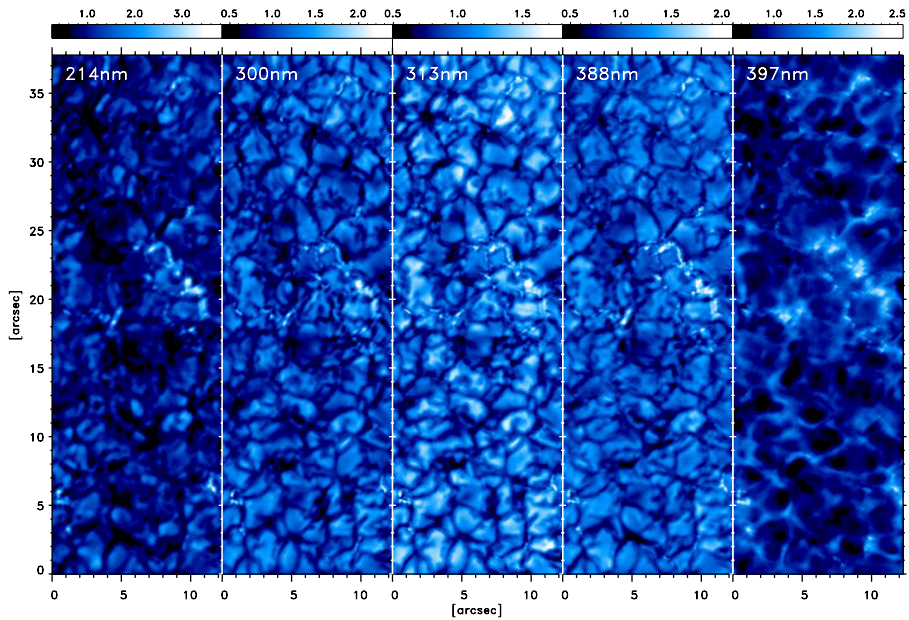}}
\caption{Phase diversity reconstructed SuFI images of the solar surface.
The field of view covers 1/20000 of the total surface of the Sun.
The depth of the false colour table is always chosen to display the full intensity range. The remarkable
constrasts in the UV wavelengths, which have been never observed before, represent an important ingredient for solar irradiance models.}
\label{data}
 \end{figure*}

A sample set of images at the different wavelengths after preliminary phase diversity reconstruction (Hirzberger {\it et al.}, 2010a) is
plotted in Figure~\ref{data}.
A careful analysis of the data (Hirzberger {\it et al.}, 2010a; Riethm\"{u}ller {\it et al.}, 2010; Solanki {\it et al.}, 2010) reveals that the observed intensity contrasts are
remarkably high, showing the
extraordinary optical performance of the total system of telescope and
instrument, especially the virtual lack of straylight contamination.

\begin{figure*}
 \centerline{\includegraphics[angle=0,width=0.5\textwidth,clip=]{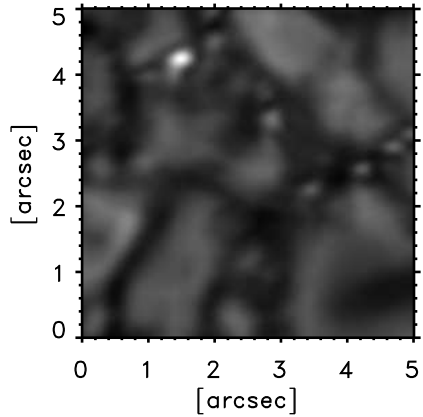}}
\caption{Detail of a phase diversity reconstructed SuFI image of the solar surface at a wavelength of 300~nm.
The depth of the false colour table is chosen to display the full intensity range.}
\label{datasmall}
 \end{figure*}

In the shortest wavelength channel at 214~nm, exposure times were one or two orders of magnitude larger than
anticipated, due to strong residual ozone extinction even in maximum flight altitude
of  37.5~km.
Therefore the ultimate spatial resolution on the solar surface could not be
reached during our science flight. This is partly due to the evolution of the
solar structures during the long exposure time; to a larger extend, however,
residual image smear on the order 0.04~arcsec (rms) at frequencies in the tens of Hertz
range corrupted the images also in the other wavelength bands, which have typical exposure times of
100~ms. For a detailed analysis of the performance of the CWS system we here
refer to Berkefeld {\it et al.} (2010).
From the phase diversity reconstruction of the wavefront in the instruments
exit pupil we have no indication of considerable optical imperfections which
could explain the limited angular resolution in our images (Hirzberger {\it et
al.}, 2010b). At 300~nm the in flight rms WFE including the telescope is typically between
$\lambda/6$ and $\lambda/5$. Since the telescope has $\lambda/12$ at 633~nm,
this implies an rms WFE of SuFI/ISliD of better than $\lambda/17$. For a
detailed analysis of the PD reconstruction and the WFE analysis we here refer
to Hirzberger {\it et al.} (2010b).

Although the CWS system could dramatically improve the effective pointing
stability of the telescope, it was not able to completely damp the
incoming disturbance spectrum of the balloon/gondola system.
The effective angular resolution of SuFI is therefore between 0.1 arcsec and 0.15 arcsec.
Although this is slightly above the theoretical diffraction limit,
the quality of the SuFI data is resounding and the data set is expected to
provide new insights in the physics of solar surface magnetism at very small
scales. ISLiD/SuFI survived the flight and the landing with only negligible
damage and are ready to be used again in {\it Sunrise}.

\begin{acknowledgements}
The German contribution to \Sunrise is funded by the Bundesministerium
f\"{u}r Wirtschaft und Technologie through Deutsches Zentrum f\"{u}r Luft-
und Raumfahrt e.V. (DLR), Grant No. 50~OU~0401, and by the
Innovationsfond of
the President of the Max Planck Society (MPG). The Spanish contribution has
been funded by the Spanish MICINN under projects ESP2006-13030-C06
and AYA2009-14105-C06 (including European FEDER funds). The HAO
contribution was partly funded through NASA grant number NNX08AH38G.
We would like to thank the anonymous reviewer and the editor for very constructive and
helpful comments on the manuscript.
\end{acknowledgements}

%%% %%%%%%%%%%%%%%%%%%%%%%%%%%%%%%%%%%%%%%%%%%%%%%%%%%%%%%%%%%%
%% Bibliography
%
% Using BibTeX
%
% \bibliographystyle{spr-mp-sola}
% %\bibliographystyle{spr-mp-sola-cnd} %% Alternative style: no title, no concluding page
% \bibliography{<bib file>}
%
% Without BibTeX
% \begin{thebibliography}{}
% \bibitem[\protect\citeauthoryear{Author}{Year}]{key}
%   <bibliographical entry>
%
% \bibitem[\protect\citeauthoryear{}{}]{}
%
%
% \end{thebibliography}

\end{article}
\end{document}